\documentclass[a4paper,12pt]{article}
\usepackage{amssymb,latexsym,amsmath}
\usepackage{subcaption}
\usepackage{fullpage}
\usepackage{mathrsfs}
\usepackage{bbold}
\usepackage{algorithm}
\usepackage{algpseudocode}
\usepackage[final]{hyperref}
\hypersetup{pdfborder={0 0 0},colorlinks=false,urlcolor=red!60!black,citecolor=blue!60!black,linkcolor=red!60!black,linktocpage=true}
\usepackage{cite}
\usepackage[pdftex]{graphicx}
\usepackage[table,dvipsnames]{xcolor}
\usepackage{multirow}
\usepackage{rotating}
\usepackage{array}
\usepackage{float}
\makeatletter \@addtoreset{equation}{section}
\makeatother

\begin{document}

\begin{titlepage}
	\thispagestyle{empty}
		
	\begin{center}

		{ \LARGE{\bf New methods for old problems: \\[4mm]
		vacua of maximal $D=7$ supergravities}}
	
		\vspace{50pt}
							
			{Dario~Partipilo$^{1,2}$}
							
			\vspace{25pt}
							
			{
				{\it  $^{1}$Dipartimento di Fisica e Astronomia ``Galileo Galilei''\\
					Universit\`a di Padova, Via Marzolo 8, 35131 Padova, Italy}
										
				\vspace{15pt}
										
				{\it   $^{2}$INFN, Sezione di Padova \\
				Via Marzolo 8, 35131 Padova, Italy}
		
			}

		\vspace{80pt}

		{ABSTRACT}
	\end{center}
	
	\vspace{10pt}
	
	Finding vacua of supergravity theories is an outstanding problem which has been tackled in several ways, and with this work we add a new method to the puzzle. We analyse the scalar sector of maximal gauged supergravity theories in seven space-time dimensions. We look for vacua of the theory by varying the embedding tensor, instead of directly minimising the scalar potential. The set of quadratic constraints arising from this procedure has been solved by means of Evolution Strategies optimisation techniques, also adopted in Artificial Intelligence studies. We develop some methods to reconstruct and obtain analytical results starting from numerical outcomes, thus obtaining the complete mass spectra. In addition to some of the known vacua, we also obtain two new Minkowski vacua.

\end{titlepage}

\baselineskip 6 mm

\allowdisplaybreaks

\date{}

\section{Introduction} 
\label{sec:introduction}
Minima of potential functions have always been of uttermost importance in physics: from the search for stable trajectories in the gravitational field of a star to the spontaneous symmetry-breaking mechanism, potentials have been shown to be very useful in understanding the physical problem at hand. With the advent of String Theory, finding minima of scalar potentials was fundamental to study the String Landscape \cite{Susskind:2003kw}, in this framework Supergravity theories proved to be a conceptual testbed for this analysis. In fact, examining the structure of the vacua for supergravities is a crucial task in order to study the patterns of supersymmetry breaking, the processes behind the generation of critical points with non-vanishing cosmological constant, and in case the latter takes negative values the possible holographic dual theory. More specifically, the supergravity regime of the bulk theory in the AdS/CFT conjecture is described
by gauged supergravities. Moreover, vacua with positive values of the cosmological constant have also recently received particular attention from the string theory community, due to the restrictions imposed upon them by the de Sitter conjecture \cite{Obied:2018sgi} and other conjectures formulated in the framework of the Swampland programme. Vacua of gauged supergravities also have the desirable feature of giving masses to scalars, thus providing a solution to the moduli stabilisation problem.\\
In this work, we will focus on maximal ($\mathcal{N}=$ 4) supersymmetric theories in $D=7$ space-time dimensions. These theories are of particular interest because of their large amount of supersymmetry that completely fixes the field content of the theory and the Lagrangian, and in light of the limited number of possible deformations to which they can be subjected. For instance, dimensional reductions of string theories or M-theory on spheres produces theories whose vacua have maximally supersymmetric holographic duals. Moreover, our interest in seven dimensions arises since it has been shown that CFTs exist up to six space-time dimensions \cite{Nahm:1977tg}, and it is possible to chart most of them once we know all the vacua of Supergravity in one more dimension. \\
Searching for minima of the scalar potentials in supergravity theories is a challenging task, and many techniques have been developed so far to tackle the problem. Progress has been achieved by pursuing three main different paths. The first consists in picking a particular theory (gaugings) and by using symmetries to truncate the field content in order to arrive at a model with a restricted number of scalars, thus making the minimisation procedure feasible. This method, introduced by N.P. Warner in \cite{Warner:1983du, Warner:1983vz} made it possible to analytically solve different minimisation conditions, thus discovering many new vacua in 4 and 5 space-time dimensions. Another proposal is to adopt a numerical approach based on machine learning libraries, and lead to the discovery of a large amount of vacua, which also allowed the analysis of mass spectra, residual supersymmetries and residual gauge groups \cite{Fischbacher:2009cj, Fischbacher:2010ki, Fischbacher:2011jx, Comsa:2019rcz, Bobev:2020ttg, Bobev:2020qev, Berman:2022jqn, Krishnan:2020sfg}. Both these techniques are based on the choice of a gauging, consequently, one has to scan one theory at a time, making them inconvenient for a complete cataloguing. Another approach has been formulated, based on the embedding tensor formalism. First discovered in the contest of the super-Higgs mechanism \cite{Li:1986tk} it has been then introduced in the framework of the analysis of supergravity vacua \cite{DallAgata:2011aa}, leading to a wealth of new results \cite{Borghese:2012qm, Borghese:2012zs, Borghese:2013dja, Catino:2013ppa, Gallerati:2014xra, Dibitetto:2011gm, Deger:2019tem, Dallagata:2021lsc} as well as providing the first example of a family of gauged theories dependent on a continuous parameter\cite{DallAgata:2012mfj}. This method does not pick a specific gauging choice, thus allowing for a simultaneous scan of different gauge theories. The last technique will produce a set of quadratic equations in $n$ variables, with $n$ the number of parameters in the embedding tensor, this problem is NP-hard, thus analytic solutions are expected to be found in very special cases, usually by restricting the number of variables, using some residual symmetry. 
In this work, we will build upon the last method and implement the scan using an evolution strategy algorithm, Covariance Matrix Adaptation (CMA-ES) \cite{hansen2006cma}, together with a Genetic Algorithm (GA). From the solutions we obtain, we can recover the analytic form of the vacua and their spectra.  

\section{Maximal supergravity in 7 dimensions} 
\label{sec:maximal_supergravity_in_7_dimensions}
We recall, in this paragraph, all the notations and properties of the tensors needed in the rest of the paper. It is well known, since the seminal paper by de Wit, Samtleben and Triggiante \cite{deWit:2002vt}, that any possible gauging of a supergravity theory is encoded in the embedding tensor $\Theta_M{}^{\alpha}$, where usually the indices $M,N,P,$ transform in the vector representation and $\alpha, \beta, etc.$ in the adjoint of the duality group (for the case of 7 dimensions $M, N, etc.$ will live in the fundamental representation $\mathbf{5}$ of SL$(5)$, so the vector representation $\mathbf{10}$ will be described by a couple of antisymmetric indices $MN=[MN]$). The job of the embedding tensor is to select the generators of the duality group, which in the case of 7 dimensions is SL(5), which will form the gauge group of the specified theory. Once this tensor is fixed, the Lagrangian and supersymmetry transformations follow, according to the analysis in \cite{deWit:2002vt}. 
We can write the new generators of the gauge group as 
\begin{equation}
X_{MN}=\Theta_{MN,P}{}^Q t^P{}_Q,
\end{equation}
where $t^P{}_Q$ are the SL(5) generators, satisfying $t^M{}_M = 0$ and the algebra
\begin{equation}
    \left[t^M{}_N, t^P{}_Q\right] = \delta^P_N t^M{}_Q -\delta^M_Qt^P{}_N .
\end{equation}
In seven space-time dimensions, the embedding tensor lives in the product of representations \textbf{10} $\otimes$ \textbf{24} = \textbf{10} + \textbf{15} + $\overline{\textbf{40}}$ + \textbf{175}. However, supersymmetry constrains it to lie in the $\textbf{15}+\overline{\textbf{40}}$, therefore it is possible to parameterize it as:

\begin{equation}
    \Theta_{MN,P}{}^{Q} = \delta_{[M}^QY_{N]P} - 2 \epsilon_{MNPRS}Z^{RS,Q} \hspace{0.3cm}.
\end{equation}
Where $Y_{MN}=Y_{(MN)}$ and $Z^{MN,P}= Z^{[MN],P}$ with $Z^{[MN,P]}=0$, that live respectiveley in the \textbf{15} and $\overline{\textbf{40}}$ representations of SL(5). 
Gauge symmetry imposes further constraints, these are in particular quadratic constraints that ensure the closure of the gauge algebra. They can be expressed in terms of the Y and Z tensors as 
\begin{equation}
    Z^{MN,P}Y_{PQ}=0 \hspace{0.3cm}, \hspace{3cm} Z^{MN,P}X_{MN}=0 \hspace{0.3cm}.
    \label{QC2a}
\end{equation}
These \textbf{15}+$\overline{\textbf{40}}$ representations can also be decomposed under the maximal compact subgroup of SL(5), that is, USp(4) $\simeq$ SO(5).
Under USp(4) they decompose as 
\begin{equation}
    \textbf{15}+\overline{\textbf{40}}\longrightarrow (\textbf{1}+\textbf{14})+(\textbf{5}+\textbf{35})\hspace{0.3cm}.
\end{equation}
Defining the scalar matrix $\mathcal{V}_M{}^{ab}$, anti-symmetric in $a$ and $b$ (a,b=1,...,4), which mediates from SL(5) to USp(4), normalised by
\begin{equation}
    \mathcal{V}_P{}^{ab}\mathcal{V}_{ab}{}^Q=\delta^Q_P, \hspace{2cm} \mathcal{V}_{ab}{}^P\mathcal{V}_P{}^{cd}=\delta^{cd}_{ab} -\frac{1}{4}\Omega_{ab}\Omega^{cd},
    \label{normalizationV}
\end{equation}
we have:
\begin{equation}
    Y_{MN}=\mathcal{V}_M{}^{ab}\mathcal{V}_N{}^{cd}Y_{ab,cd} \hspace{0.1cm}, \hspace{3cm} Z^{MN,P}=\sqrt{2}\mathcal{V}_{ab}{}^M \mathcal{V}_{cd}{}^N \mathcal{V}_{ef}{}^P \Omega^{bd}Z^{(ac)[ef]}.
\end{equation}
We call B, $B^{[ab]}{}_{[cd]}$, $C_{[ab]}$ and $C^{[ab]}{}_{(cd)}$ the irreducible components that live in \textbf{1}, \textbf{14}, \textbf{5} and \textbf{35} of USp(4), respectively. 
In particular, they satisfy
\begin{align}
    \begin{split}
        C^{ab}&=C^{[ab]}, \hspace{1.2cm} \Omega_{ab}C^{ab}=0,\\
        B^{ab}{}_{cd} &= B^{[ab]}{}_{[cd]}, \hspace{1.05cm} B^{ab}{}_{cb}=0, \hspace{1cm} \Omega_{ab}B^{ab}{}_{cd} = 0 = \Omega^{ab}B^{cd}{}_{ab}, \\
        C^{ab}{}_{cd} &= C^{[ab]}{}_{(cd)}, \hspace{1cm} C^{ab}{}_{cb} = 0, \hspace{1cm} \Omega_{ab}C^{ab}{}_{cd}=0.
    \end{split}
    \label{TtensRep}
\end{align}
In terms of these, the Y and Z tensors become
\begin{align}
    Y_{ab,cd}&= \frac{1}{\sqrt{2}}\left[(\Omega_{[a|[c|}\Omega_{b]d]}-\frac{1}{4}\Omega_{ab}\Omega_{cd})B + \Omega_{[a|e|}\Omega_{b]f}B^{[ef]}{}_{[cd]}\right]\hspace{0.3cm},\\
    Z^{(ab)[cd]}&= \frac{1}{16}\Omega^{a[c}C^{d]b}+\frac{1}{16}\Omega^{b[c}C^{d]a}-\frac{1}{8}\Omega^{ae}\Omega^{bf}C^{cd}{}_{ef}\hspace{0.3cm}.
\end{align}
Mapping the $\Theta$-tensor from SL(5) to USp(4) we get the T-tensor defined as: 
\begin{equation}\label{Ttens}
   T_{(ef)[ab]}{}^{[cd]} = \sqrt{2}\Omega^{h[c}\delta^{d]}_{(e}\mathcal{V}^M{}_{f)h}\mathcal{V}^N{}_{ab}Y_{MN}-2\sqrt{2}\epsilon_{MNPQR}Z^{PQ,S}\mathcal{V}^M{}_{eg}\mathcal{V}^N{}_{fh}\mathcal{V}_{ab}\,^R\mathcal{V}_S{}^{cd}\Omega^{gh}\hspace{0.3cm}.
\end{equation}
In terms of these new tensors, it is possible to rewrite the quadratic constraints as follows.
\begin{align}\label{constraints}
    \begin{split}
    Z^{(ab)[ef]}&\left[\Omega_{ce}\Omega_{df}B+\Omega_{eg}\Omega_{fh}B^{[gh]}{}_{[cd]}\right]=0\hspace{0.3cm},\\
    &Z^{(ab)[cd]}T_{(ab)ef}{}^{gh}=0\hspace{0.3cm}.
    \end{split}
\end{align}
\pagebreak
Let us now come to the core of our analysis: the scalar potential and the mass matrices. As usually happens in gauged supergravities, the scalar potential can be expressed as the difference of the squared fermion shifts, the latter are defined to be:
\begin{align}
    A_1^{ab}&\equiv -\frac{1}{4\sqrt{2}}\left(\frac{1}{4}B\Omega^{ab}+\frac{1}{5}C^{ab}\right)\hspace{0.3cm},\\
    A_2^{d,abc}&\equiv\frac{1}{2\sqrt{2}}\left[\Omega^{ec}\Omega^{fd}\left(C^{ab}{}_{ef}-B^{ab}{}_{ef}\right)+\frac{1}{4}\left(C^{ab}\Omega^{cd}+\frac{1}{5}\Omega^{ab}C^{cd}+\frac{4}{5}\Omega^{c[a}C^{b]d}\right)\right]\hspace{0.3cm}.
    \label{fermionshifts}
\end{align}
Then, the scalar potential is given by:
\begin{equation}
    V=\frac{1}{8}|A_2|^2 -15|A_1|^2 =-\frac{1}{128}\left(15B^2 +2C^{ab}C_{ab}-2B^{ab}{}_{cd}B^{cd}{}_{ab}-2C^{[ab]}{}_{(cd)}C_{[ab]}{}^{(cd)}\right)\hspace{0.1cm}.
    \label{potential7D}
\end{equation}
From this we can compute the variation of the scalar potential under a variation of the scalar fields given by $\delta_{\Sigma}\mathcal{V}_M{}^{ab}=\Sigma^{ab}{}_{cd}\mathcal{V}_M{}^{cd}$, where $\Sigma^{ab}{}_{cd}$ is a variation along the scalar manifold SL$(5)/$USp$(4)$, living in the \textbf{14} representation.
\begin{align}\label{eom}
    \begin{split}
        \delta_{\Sigma}V=&-\frac{1}{16}B^{[ab]}{}_{[cd]}B^{[cd]}{}_{[ef]}\Sigma^{[ef]}{}_{[ab]}+\frac{1}{32}BB^{[ab]}{}_{[cd]}\Sigma^{[cd]}{}_{[ab]}-\frac{1}{64}C^{[ab]}C_{[cd]}\Sigma^{[cd]}{}_{[ab]}\\
        &+\frac{1}{32}C^{[ab]}{}_{(ef)}C_{[cd]}{}^{(ef)}\Sigma^{[cd]}{}_{[ab]}-\frac{1}{8}C^{[ce]}{}_{(af)}C^{[df]}{}_{(be)}\Sigma^{[ab]}{}_{[cd]}\hspace{0.3cm}.
    \end{split}
\end{align}
In order to find the vacua of the theory, we need to solve this equation of motion together with the quadratic constraints (\ref{constraints}) or (\ref{QC2a}). 
Once we obtained the solutions to these quadratic equations, we studied the gauging, the residual gauge group, the residual amount of supersymmetry preserved by each vacuum, the mass spectra, and the values of the cosmological constant.

On the vacuum, residual supersymmetry is given by the number of non-vanishing $\epsilon^a$ parameters satisfying 
\begin{equation}\label{residualsusy}
    A_{2a,bcd}\epsilon^a=0\hspace{0.3cm}.
\end{equation}
On the other hand, for what concerns the masses, we have computed the various mass matrices. For the gravitinos, we have:
\begin{equation}
    \mathcal{M}_{ij}^{3/2}=\frac{5}{2}A_{1ij}\hspace{0.3cm}.
\end{equation}
The mass matrix for the fermions is given by 
\begin{equation}
    \mathcal{L}_{m-1/2}=\frac{1}{4\sqrt{2}}\left[\frac{3}{32}\delta^b_d\delta^c_e B +\frac{1}{8}\delta^b_d\Omega_{ef}C^{fc} +B^{bc}\,_{de}-C^{bc}\,_{de}\right]\overline{\chi}_{abc}\chi^{ade},
    \label{fermionbaremass}
\end{equation}
in order to compute the masses for the fermions, we must recall that $\chi^{abc}$ transforms in the $\mathbf{16}$ of Usp$(4)$
\begin{equation}
    \chi^{abc}=\chi^{[ab]c}, \hspace{1cm} \Omega_{ab}\chi^{abc}=0, \hspace{1cm} \chi^{[abc]}=0,
\end{equation}
and therefore also the mass matrix must lie in the $\mathbf{16}\times \mathbf{16}$ representation of Usp$(4)$.
The vector mass matrix is contained in the scalar kinetic term $\frac{1}{2}P_{\mu ab}{}^{cd}P^{\mu}_{cd}{}^{ab}$, where $P_{\mu ab}{}^{cd}$ is defined by the gauge covariant space-time derivative of the scalar fields
\begin{equation}
    \mathcal{V}_{ab}{}^M (\partial_{\mu}\mathcal{V}_M{}^{cd} -g A_{\mu}^{PQ}X_{PQ,M}{}^N \mathcal{V}{}^{cd})\equiv P_{\mu ab}{}^{cd} +2 Q_{\mu [a}{}^{[c}\delta_{b]}^{d]}.
\end{equation}
The $P_{\mu ab}{}^{cd}$ lies in the $\mathbf{14}$ representation of $\mathfrak{usp}$(4), while $Q_{\mu a}{}^{c}$ in the $\mathbf{10}$, (these must be imposed before computing the masses).
Analogously, the mass term for the 2-forms arises from the kinetic term of the vectors, namely $\Omega_{ac}\Omega_{bd}\mathcal{H}_{\mu \nu}{}^{ab}\mathcal{H}^{cd\mu\nu}$, where $\mathcal{H}_{\mu \nu}{}^{ab}$ is given by the modified field strength tensor:
\begin{equation}
    \mathcal{H}^{PQ}_{\mu\nu}=\mathcal{F}_{\mu\nu}^{PQ} +gZ^{PQ,R}B_{\mu\nu R},
    \label{vecfieldStrength}
\end{equation}
Similarly, the mass term for the 3-forms, $S^N_{\mu\nu\rho}$, arises from the kinetic term of the 2-forms, $\Omega^{ac}\Omega^{bd}\mathcal{H}_{\mu\nu\rho ab}\mathcal{H}^{\mu\nu\rho}_{cd}$. 
The covariant field strength for the 2-forms is given by:
\begin{equation*}
    \mathcal{H}_{\mu\nu\rho M} =3D_{[\mu}B_{\nu\rho]M}+6\epsilon_{MNPQR}A^{NP}_{[\mu}\bigg(\partial_{\nu}A_{\rho}^{QR} +\frac{2}{3}g X_{ST,U}{}^Q A_{\nu}^{RU}A_{\rho]}^{ST}\bigg) +g Y_{MN}S^N_{\mu\nu\rho} .
\end{equation*}
Note that the kinetic term for the 3-forms 
\begin{equation}
    e^{-1}\mathcal{L}_{kin }=-\frac{1}{9}\epsilon^{\mu\nu\rho\lambda\sigma\tau\kappa}gY_{MN}S^M_{\mu\nu\rho}D_{\lambda}S^N_{\sigma\tau\kappa}
\end{equation}
is linear in the derivative, so what one really obtains from this procedure is the mass matrix, not the square mass matrix. Furthermore, this kinetic term is not canonically normalised, in fact it is in the schematic form $Y_{MN} S^M D S^N$, so the true masses are obtained once one multiplies the mass matrix that arises from the kinetic term of the 2-forms by $Y_{MN}^{-1}$.
For the scalar masses, we note that it is possible to parameterise the scalar fields in terms of the USp$(4)$-invariant, symmetric unimodular matrix $\mathcal{M}_{MN}$ defined by
\begin{equation}
    \mathcal{M}_{MN} \equiv \mathcal{V}_M\,^{ab}\mathcal{V}_N\,^{cd}\Omega_{ac}\Omega_{bd}.
\end{equation}
The scalar potential, written in terms of $\mathcal{M}_{MN}$, is 
\begin{align}
   \begin{split}
       V=&\frac{1}{64}\left(2\mathcal{M}^{MN}Y_{NP}\mathcal{M}^{PQ}Y_{QM} -\left(\mathcal{M}^{MN}Y_{MN}\right)^2\right)\\&+Z^{MN,P}Z^{QR,S}\left(\mathcal{M}_{MQ}\mathcal{M}_{NR}\mathcal{M}_{PS}-\mathcal{M}_{MQ}\mathcal{M}_{NP}\mathcal{M}_{RS}\right).
   \end{split} 
\end{align}
Therefore, by means of $\delta_{\Sigma}\mathcal{V}_M{}^{ab}=\Sigma^{ab}{}_{cd}\mathcal{V}_M{}^{cd}$, it is possible to calculate the second variation of the potential and, therefore, the scalar mass matrix, always recalling that $\Sigma^{ab}{}_{cd}$ belongs to the $\mathbf{14}$ representation of USp$(4)$.
Eq. (\ref{normalizationV}) has been widely used to calculate the masses. 
A complete study of the field content, Lagrangian, and symmetries for maximal supersymmetric gauged supergravities in 7 space-time dimensions is given in \cite{Samtleben:2005bp}.

\section{Scanning for vacua: methodologies}
\begin{figure}[ht]
	\begin{center}
        \includegraphics[scale=.3]{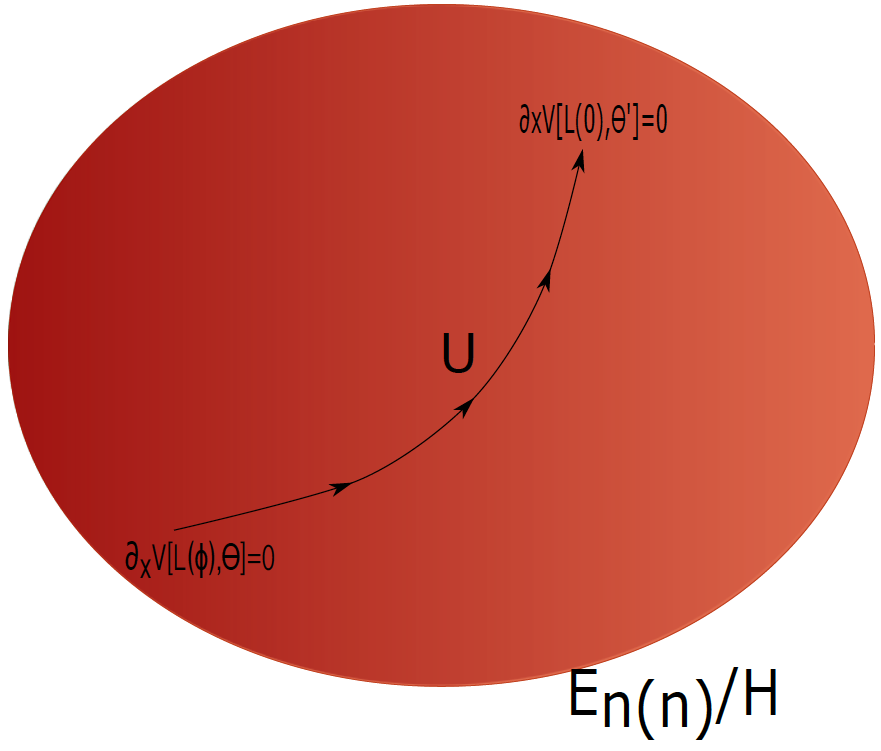} 
 	\end{center}
\caption{E$_{n(n)}$ transformations (U) mapping any point to the origin of homogeneous space E$_{n(n)}/H$}
\label{Homtransf}
\end{figure}

\begin{figure}[H]
\begin{subfigure}{.52\textwidth}
  \centering
  \includegraphics[width=.7\linewidth]{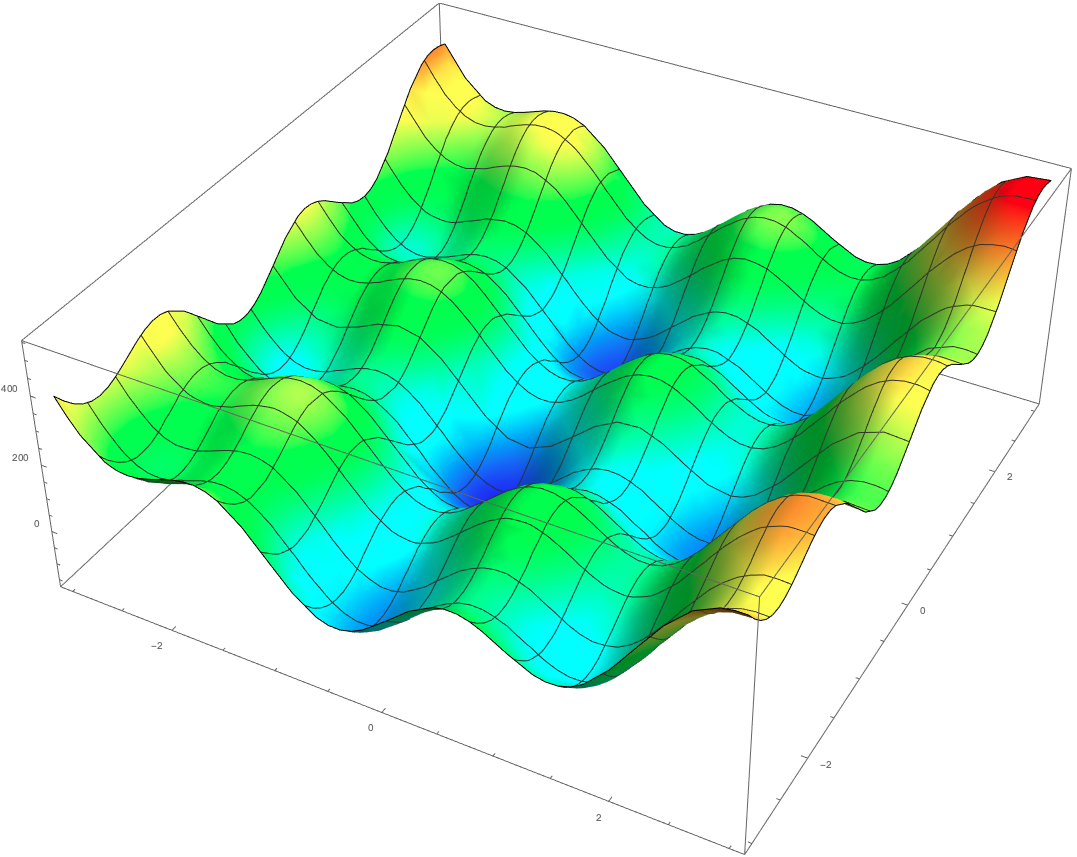}
  \caption{Example of fitness function $F_{fit}$ without modification}
  \label{fig:senzagaussiana}
\end{subfigure}%
\hspace{0.2cm}
\begin{subfigure}{.52\textwidth}
  \centering
  \includegraphics[width=.7\linewidth]{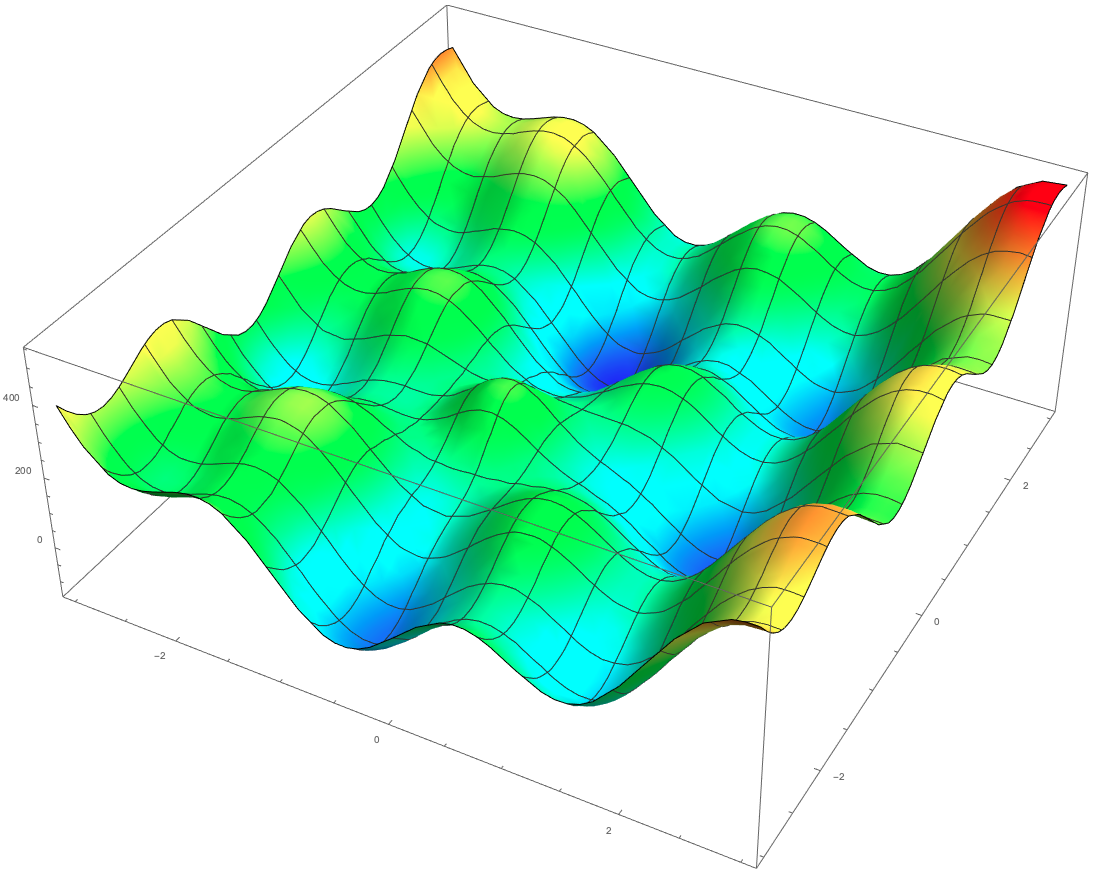}
  \caption{Example of fitness function with a \\ Gaussian modification}
  \label{fig:congaussiana}
\end{subfigure}
\caption{Modification of the fitness function after the algorithm has found a minimum.}
\label{fig:gaussiana}
\end{figure}

\begin{figure}[H]
    \centering
    \subfloat[Uncorrelated Variables]{
        \label{fig:uncorrelated}
        \includegraphics[width=.5\linewidth]{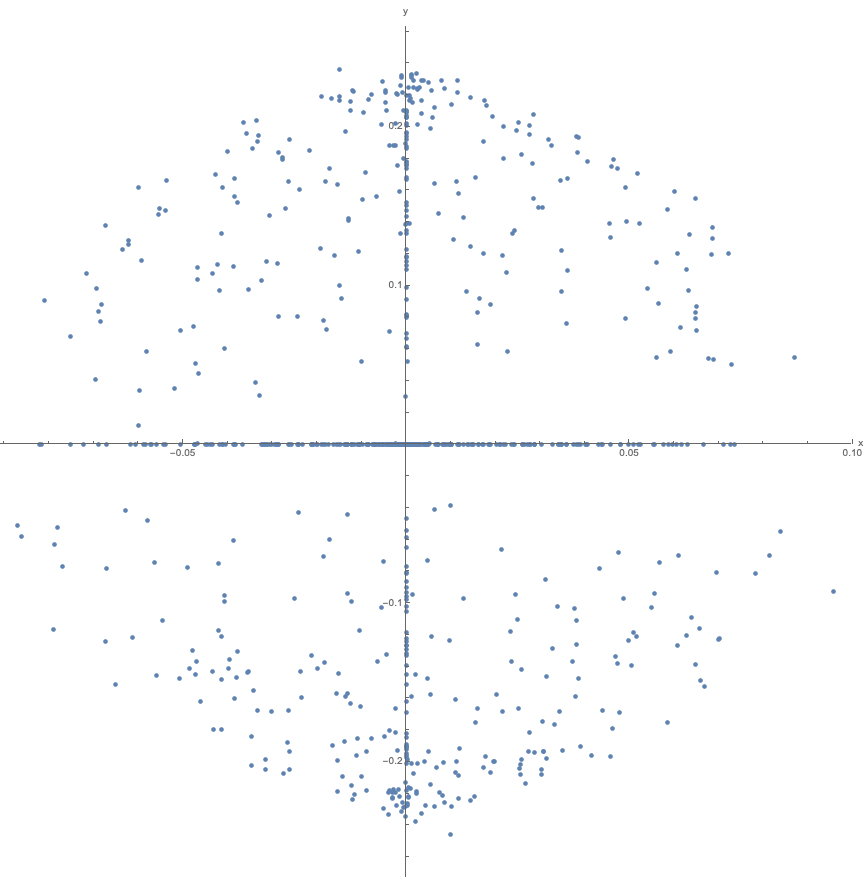}
    }
    \subfloat[No dependance among the variables]{
        \label{fig:0yaxis}
        \includegraphics[width=.5\linewidth]{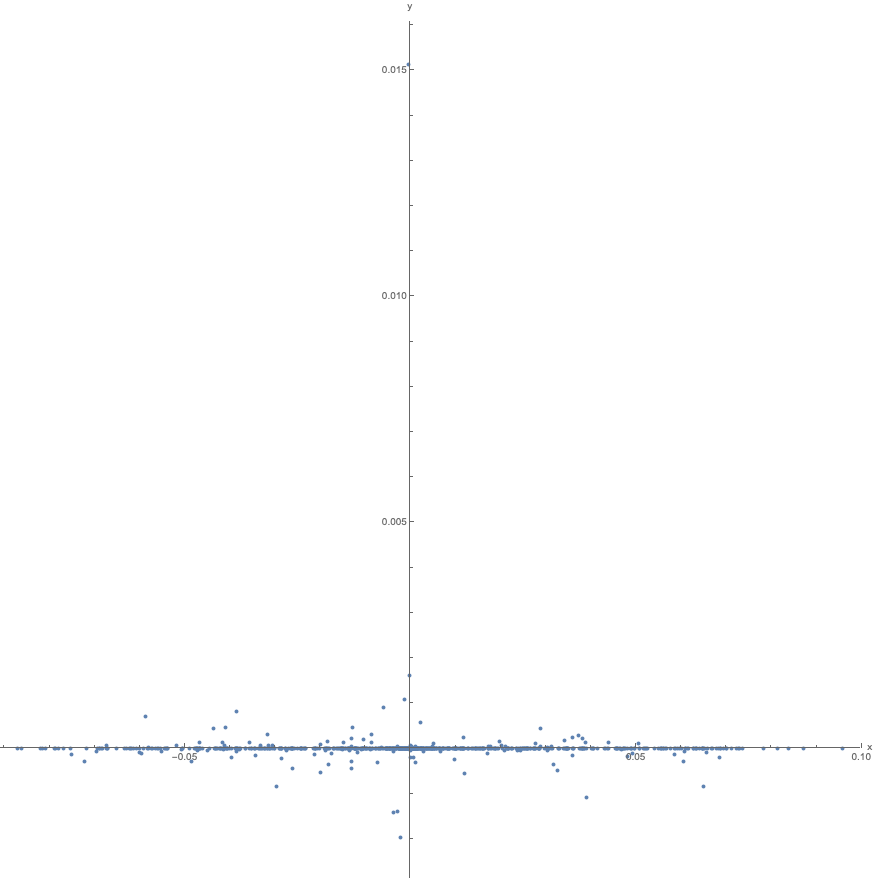}
    }
    \\
    \subfloat[variables linearly related]{
        \label{fig:xycorrelated}
        \includegraphics[width=.5\linewidth]{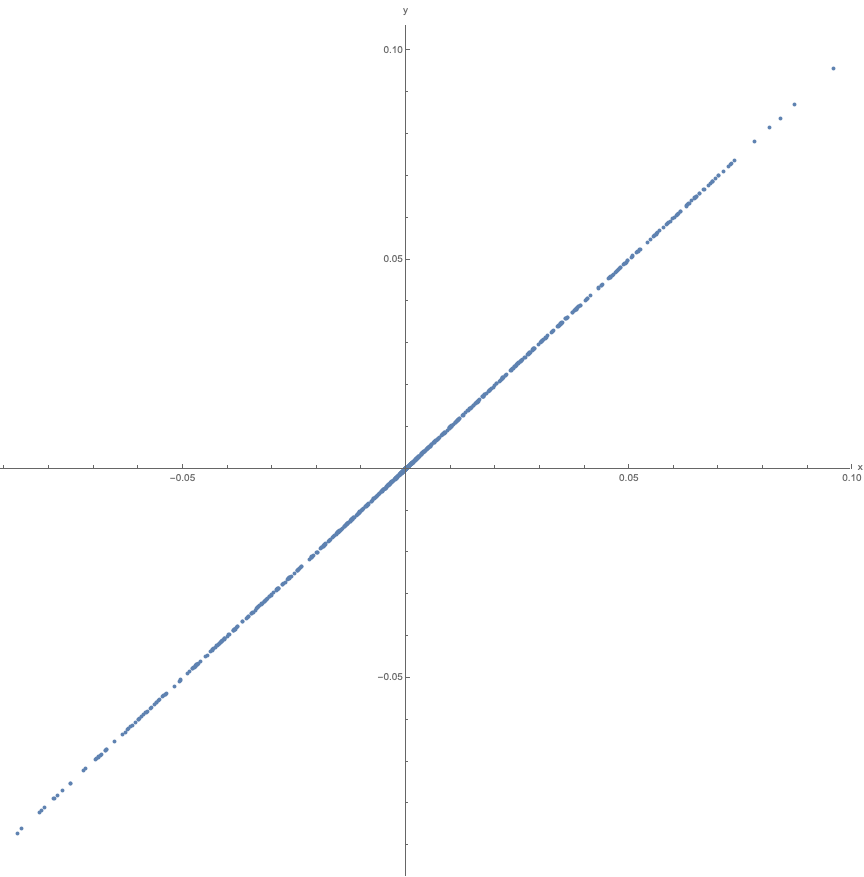}
    }
    \subfloat[Galois couple]{
        \label{fig:doublecorrelation}
        \includegraphics[width=.5\linewidth]{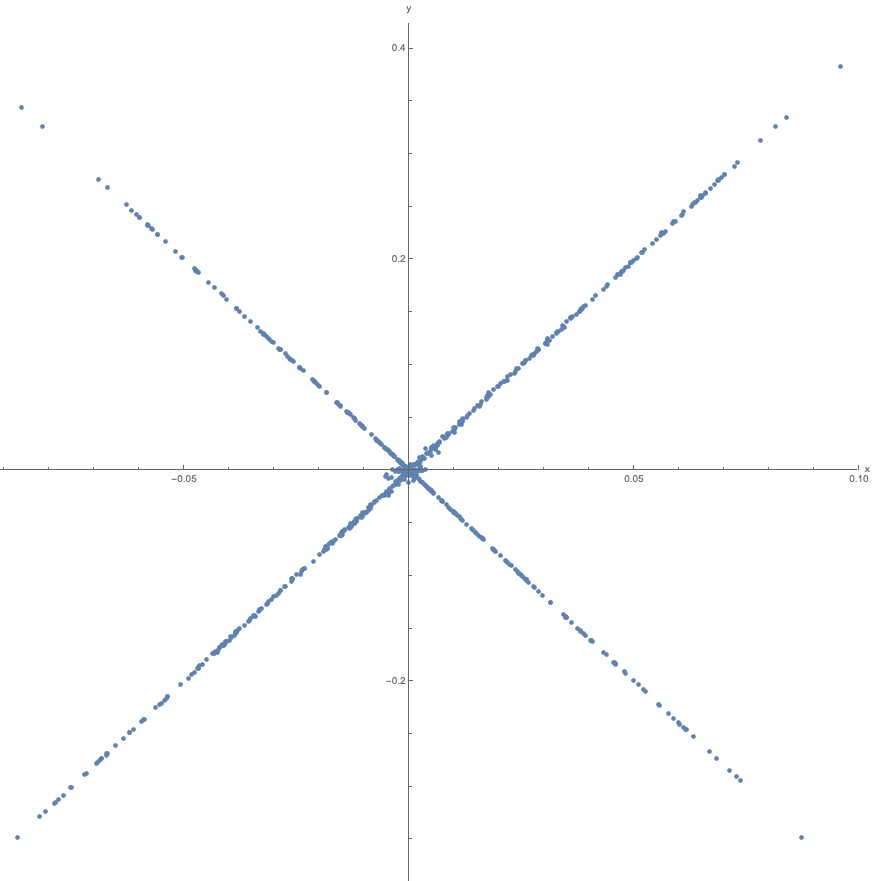}
    }
    \caption{Correlation among the variables}
    \label{fig:correlation}
\end{figure}

The approach we propose to chart the vacua of the theory is based on optimisation techniques, some of the methods we adopted are not unknown to the string theory community. Indeed, we used a combination of analytical and numerical tools; our starting point will be the analysis performed in \cite{DallAgata:2011aa}. There, the authors showed how every vacuum found in an explicit gauging of the theory can be mapped, by means of a duality transformation, to the origin of the scalar manifold, as shown in Fig. \ref{Homtransf}. This procedure is allowed by the homogeneous nature of the coset manifold, which in the case of maximal supergravities is of the form E$_{n(n)}/H$ (we do not include the Trombone symmetry in our analysis), with $n$ the number of internal dimension and $H$ the maximal compact subgroup of E$_{n(n)}$.
Due to this method, we will be able to scan the entire space of all gaugings at once. At the origin, the physical scalars are all set to 0, but the parameters of the embedding tensor are not fixed, so in addition to considering the minimisation conditions $\delta_{\Sigma}V=0$, with $\delta_{\Sigma}V$ given in eq. (\ref{eom}), we must also impose the quadratic constraint, eq. (\ref{QC2a}) or (\ref{constraints}), (the linear constraint has been imposed when choosing to work with the representations $(\mathbf{1} + \mathbf{14}) + (\mathbf{5} + \mathbf{35})$ in (\ref{TtensRep})). The maximal compact subgroup of SL$(5)$, SO$(5)$, can be used to remove part of the variables, in this case 10, from the global system. In particular, we can always use local SO$(5)$ to diagonalize $Y_{MN}$. 
This leads to the problem of solving systems of multivariate quadratic equations (($\mathcal{MQ}$) problem), which is known to be a $\mathcal{NP}$ complete task. Indeed, various cryptosystems relies on this problem in some way, $\mathcal{MQ}$ is also considered a possibility for potentially post-quantum cryptosystems, namely, it is believed not to be solved by quantum algorithms as well. Therefore, we introduce a different, related, problem: we are going to solve the systems with a certain approximation, and this can always be done in polynomial time. Indeed, optimisation techniques have been widely used by physicists and mathematicians to find solutions to systems of equations. In order to transform our problem, recalling that we have a system of homogeneous equations of the form $\sum_i^n d_i =0$, we introduce a fit function $f_{fit}$ defined as
\begin{equation}
    f_{fit}=\sum_i^n (d_i)^2
\end{equation}
where $i$ runs over the length of the system and $d_i$ represents the polynomials in the system. Obviously, the global minima of the fit function correspond to $f_{fit}=0$ and are the solutions to our system. This also implies that verifying that a minimisation point is also a solution takes no time. The first approach invented to solve a set of equations simultaneously is the Gradient Descent algorithm (GD), created by A. Cauchy, and since then many other attempts have been tried. Genetic Algorithms (GA), which are an instance of the broad class of Evolutionary Strategies algorithms, have already been widely used by theoretical physicists \cite{Abel:2018ekz, Abel:2014xta, Yamaguchi:1999hq, Markham:1998sf, Allanach:2004my}, they have the advantage to work with ill-behaved functions, while methods such as GD or stochastich GD work only on differentiable fitness functions. In this work, we are going to introduce yet another method in the set of tools used for the research of supergravity vacua, the Covariance Matrix Adaptation Evolutionary Strategy (CMA-ES). In our search of minima of the scalar potential, we used a combination of GAs and CMA-ES, we will not dive into the explanation of GAs, since they are already known by the community, a nice introduction about these topics and other Data Science applications to String Theory can be found in \cite{Ruehle:2020jrk}. Instead, we present the new technique, based on CMA-ES.
CMA is an Evolutionary Strategy algorithm, therefore, it consists of a series of steps (generations), at each of them some operations are carried out before going to the next step. CMA is based on the adaptation of the covariance matrix associated with the multivariate normal distribution, from which we draw a new set of candidate solution points at each generation. 
The covariance matrix is not the only object that needs to be adapted at each step; the procedure implies the computation of the new mean and the overall step size as well. To compute the mean, we order the candidate solution points, based on the fitness function, and select the best $\mu$, which results in a higher ``evolution pressure''. On the other hand, the Covariance Matrix is adapted in order to maximise the likelihood of precedent successful steps. Thus, CMA performs an iterated analysis of the principal components at every generation, while retaining all the principal axes. A deeper analysis of the algorithm is presented in Appendix \ref{sec:CMA}. Let us now describe some details of the analisis. Parallelisation proved to be fundamental in making the process faster and allowing for a more global search scan. Indeed, with CMA-ES we need to choose a starting mean $\mathbf{m}^0$ and step size $\mathbf{\sigma}^0$, together with some hyperparameters such as the time allowed to carry the computations (`timeout' parameter), the minimum value accepted for the fitness function in order to declare that a minimum was achieved (`ftarget' parameter), the precision of this result (`tolfun'), the population size $\lambda$ and whether or not to activate elitist research. Parallelisation gave us the opportunity to choose more initial means $\mathbf{m}^0$ at a time, thus scanning a larger area of the parameter space. Some useful techniques have been used to adapt the algorithms to our specific problem and to render the analysis of numerical results faster. \\
First of all, we are dealing with systems of homogeneous quadratic equations, therefore, their solutions always pass from the origin, which helps us to restrict the area of research when setting the initial mean for the multivariate normal distribution. On the other hand, we must pay attention because the origin of the reference system is a trivial solution for any homogenous system of equations and, therefore, starting close to it may lead us always there. This could be avoided by modifying the fitness function.  We found the following definition to be very efficient:
\begin{equation}
    F_{fit}=\begin{cases}
    f_{fit} \hspace{1cm} \text{for} \hspace{0.5cm} \sum_i^n x_i^2 > \text{threshold},\\ 
    10000 \hspace{1cm} \text{for} \hspace{0.5cm} \sum_i^n x_i^2 \le \text{threshold}.
    \end{cases}
\end{equation}
Basically, we created a step function to avoid the algorithm from always converge in $\mathbf{0}$. The threshold must be chosen in such a way as to leave enough parameter space near the origin to complete a full scan without hitting the barrier too often. A technique has also been implemented to avoid the algorithm from returning to the previously found minima. It consisted of adding a multivariate normal distribution function centred on the minima on top of the fitness function, as illustrated in Fig. \ref{fig:gaussiana}. Basically what one does is to add a series of umbrellas on top of the fitness function to stop the algorithm from going in those directions already analised.
In our case, though, the solutions to the systems are manifolds in the parameter space, so an infinite number of umbrellas would be needed to prevent the algorithm from finding again the vacuum structure under consideration. Imagine having a straight line in 2-dimensions and starting to cover it with 2-dimensional multivariate normal distributions. In 7 space-time dimensions, we were interested, above all, in (Anti)-de Sitter vacua, and this information can be used to further simplify the numerical search. Indeed, taking into account the potential in (\ref{potential7D}) written in terms of the fermion shifts (\ref{fermionshifts}), we can see that the potential is written as the difference between 2 squared terms. For AdS we want to impose $V=-k$, with $k$ constant, we can always normalise $k$ to be 1, so we need to add another equation (quadratic) to our system, analogously for dS we need to solve $V=1$. We can always make a change of variables to completely solve this new constraint, for example, considering the case of AdS vacua, calling $z_i$ the variables contained in $A_2^{d,abc}$ and $x_i$ the ones in $A_1^{ab}$:
\begin{equation}
    \begin{cases}
        z_i \rightarrow& \frac{\sinh[\psi] w_i}{\sqrt{\sum_j w_j^2}},\\
        x_i \rightarrow& \frac{\cosh[\psi] u_i}{\sqrt{\sum_j u_j^2}}.
    \end{cases}
    \label{changevar}
\end{equation}
This change of variables introduces one more variable, $\psi$, but solves the constraint $V=-1$, thus removing a bunch of solutions from the system. By doing so, we will be certain that the algorithm will look only for vacua with a negative cosmological constant and that the other vacua disappeared from its landscape. Analogously, we can solve the constraint $V=1$ for dS vacua by exchanging $sinh$ with $cosh$ in eq. (\ref{changevar}). For the Minkowski vacua we just add one more homogeneous quadratic equation ($V=0$) to the system. The initial points have been chosen as follows, for what concerns the variables $w_i$ and $u_i$ in (\ref{changevar}), we draw them from a normal distribution with mean $\mathbf{0}$ and standard deviation $\sigma = 4$, instead, the starting points for $\psi$ have been evenly displaced in an interval from 0.1 to 0.5 (note that the initial variables scale exponentially with $\psi$, so there is no need to reach high values for the latter). We used GA techniques, such as mutation and crossing over, when the algorithm CMA got stuck in a local minimum, in order to add some noise to the candidate solution points and move them away from the local minima.
Let us now describe the analysis of the results.
First, we remove the solution points that are close to each other, up to a certain threshold that we set to be equal to $3 \sigma$, thus 3 times the step size, removing all candidate solutions corresponding to vacua already present in the set of solutions. Then, we studied the residual amount of supersymmetry, the gravitini masses, the signature of the Cartan matrix (providing information about the number of compact and non-compact generators of the gauge group of the theory), and the rank of $X_{PQ}{}^R$ giving us the dimension of the gauge group. The latter will provide us with everything that is needed to understand what gauging the vacuum belongs to. With all this information, it is possible to group the solution points and extract some very useful relations among them. Indeed, by plotting a variable $x_i$ or $z_i$, against all other variables, it has been possible to reconstruct some analytical relations, as shown in Fig. (\ref{fig:correlation}), even by means of linear fitting. This is mainly possible when the number of unknowns is small, on the other hand, once we know the residual symmetry group of the solution, we can always use the technique of restricting the scan only to those variables which are present when that particular residual gauge symmetry has been imposed. This can lead to a new system that may be solved completely analytically, or we can always iterate the previous steps. The pseudocode for the search for AdS vacua is reported in Algorithm \ref{alg:cap}.

\begin{algorithm}[H]
\caption{Scanning for AdS Vacua}\label{alg:cap}
\Comment{Initialize the number of variables, the population size, the number of values for $\psi$ and the Step-size for CMA-ES}
\\
\hspace*{\algorithmicindent}\textbf{Input} $\# \, variables$ , popSize , psiSize , StepSize 
\\
\begin{algorithmic}
\State  $ \overline{u}_i , \overline{w}_i \gets  \mathcal{N}(\mu = 0,\,\sigma=3 , \, size=(popSize,\, \# \, variables))$
\State $\overline{\psi}_i \gets [0.1,\, 1,\, size = psiSize]$ \Comment{Linearly spaced array between 0.1 and 1}
\\
\Function{$f_{fit}$}{$u_i, w_i, \psi_i$}
    \State\Return $\sum_j p_j(u_i,w_i, \psi_i)^2$   
\\ \Comment{Sum of squared polynomials with the variables defined as in \ref{changevar}}
\EndFunction
\Function{$CMA\_minimization$}{$function, \overline{u}_i, \overline{w}_i, \overline{\psi}_i$, StepSize}
    \State\Return Candidate Solution Points in $\mathbb{R}^n$   
\\ \Comment{Hyperparameters, such as 'ftarget', 'seed', 'timeout', etc., must be set as well}
\EndFunction
\\
\State $Good\_\overline{u}_i$, $Good\_\overline{w}_i \gets \{\}$
\While{$j \le psiSize$}
    \If{$f_{fit}(\overline{u}_i, \overline{w}_i, \overline{\psi}_j) \le threshold$ }
        \State $Good\_\overline{u}_i.insert(\overline{u}_i)$
        \State $Good\_\overline{w}_i.insert(\overline{w}_i)$\\
        \Comment{Accept only starting points with a fitness function lower than a certain threshold}
    \EndIf
    \State $Candidate\_Solution \gets CMA\_minimization(f_{fit}, Good\_\overline{u}_i, Good\_\overline{w}_i, \overline{\psi}_j, StepSize)$
    \If{$Candidate\_Solution$ is a local minimum}
        \State Use GA techniques
    \EndIf
\EndWhile\label{CMA}
\For{\texttt{$<k=1,\, k++,\, k\le Length(Candidate\_Solution)>$}}
    \For{\texttt{$<j= k+1,\, j++,\, j\le Length(Candidate\_Solution)>$}}
        \If{$|Candidate\_Solution[k]-Candidate\_Solution[j]|^2 \le 3\cdot StepSize$}
        \State $Candidate\_Solution.remove(j)$
        \EndIf
    \EndFor
\EndFor
\State \Return Candidate\_Solution
\end{algorithmic}
\end{algorithm}
\pagebreak
\section{Results}
In this section we are going to present the vacua found through the methods presented before. As has been explained, we have been able to reconstruct directly or obtain analytical solutions from residual symmetry considerations starting from numerical ones. We found the two well-known AdS vacua, the maximal supersymmetric one \cite{Pilch:1984xy} and the non-supersymmetric one \cite{Pernici:1984zw}, of which we present the mass spectrum. In addition, we found 2 new Minkowski vacua. The summary of the vacua is presented in Table \ref{summary7d}.

\begin{table}[H]
\centering
\rowcolors{1}{white}{gray!15}
\begin{tabular}{|c|cccc|}
\hline
vacuum & susy & G$_{gauge}$ & G$_{res}$ & ref.\\[2mm]
\hline\hline
A1 & 4 & SO(5) & SO(5) & \cite{Pilch:1984xy}\\[2mm]
A2 & 0 & SO(5) & SO(4) & \cite{Pernici:1984zw}\\[2mm]
M1 & 0 & U(1)$ \ltimes {\mathbb R}^{4}$ & U(1) & here\\[2mm]
M2 & 0 & U(1)$ \ltimes {\mathbb R}^{6}$ & U(1) & here\\[2mm]
\hline
\end{tabular}
\caption{Summary of vacua}
\label{summary7d}
\end{table}
All AdS vacua appear in the SO(5) theory, which is a consistent truncation of M-theory compactified on $S^4$.

\subsection{Mass spectra} 
\label{sub:mass spectra}
In this section we present some of the masses for all the vacua listed in table \ref{summary7d}. The masses for the AdS backgrounds are normalised in terms of the squared AdS radius $L^2 = |15/V|$, so that supersymmetric gravitinos have a normalised squared mass of 25/16.

\hspace{-1.2cm}
\begin{tabular}{cc}
\begin{minipage}{.48\linewidth}
\begin{table}[H]
\rowcolors{1}{white}{gray!15}
\begin{tabular}{|c|c|}
	\hline
	$L^2 m^2_{3/2}$ & $\left[\frac{25}{16}\right]_{\times 4}$  \\[2mm]
	$L^2 m^2_{vec}$ & $[0]_{\times 10}$  \\[2mm]
	$L^2 m^2_{2-forms}$ &   $[0]_{\times 5}$  \\[2mm]
	$L^2 m^2_{3-forms}$ &   $[1]_{\times 5}$  \\[2mm]
	$L^2 m^2_{1/2}$ & $\left[\frac{9}{16}\right]_{\times 16}$ \\[2mm]
	$L^2 m^2_{scal}$ & $[8]_{\times 14}$ \\[2mm]
	\hline
\end{tabular}
\centering
\caption{Masses for the AdS vacuum A1}
\end{table}
\end{minipage}
&
\begin{minipage}{.48\linewidth}
\begin{table}[H]
\rowcolors{1}{white}{gray!15}
\begin{tabular}{|c|c|}
	\hline
	$L^2 m^2_{3/2}$ & $\left[\frac{27}{16}\right]_{\times 4}$  \\[2mm]
	$L^2 m^2_{vec}$ & $\left[\frac{3}{4}\right]_{\times 4}$, $\left[0\right]_{\times 6}$  \\[2mm]
	$L^2 m^2_{2-forms}$ &   $[0]_{\times 5}$  \\[2mm]
	$L^2 m^2_{3-forms}$ &   $\left[\frac{3}{4}\right]_{\times 4}$, $3_{\times 1}$  \\[2mm]
	$L^2 m^2_{1/2}$ & $\left[0\right]_{\times 4}$,  $\left[\frac{3}{16}\right]_{\times 12}$\\[2mm]
	$L^2 m^2_{scal}$ & $\left[-12\right]_{\times 1}$, $[12]_{\times 9}$,  $[0]_{\times 4}$ \\[2mm]
	\hline
\end{tabular}
\centering
\caption{Masses for the AdS vacuum A2}
\end{table}
\end{minipage}
\end{tabular}
\\
\bigskip
\\
The spectrum for the vacuum A1 agrees with the one found in \cite{Pilch:1984xy}, and it is determined by the representation of the superconformal algebra in six dimensions \cite{Gunaydin:1984wc, Gunaydin:1999ci}.
The scalar masses for the A2 case are the same as those in \cite{Pernici:1984zw}, ensuring that this is the same non-supersymmetric AdS vacuum.
In order to compute the masses for the fermions, we note that in the A2 case supersymmetry is completely broken, therefore a super-Higgs mechanism is in action there. Consequently, 4 fermions, which correspond to the Goldstinos, must be eaten by the corrisponding gravitinos. The relevant part of the Lagrangian is 
\begin{equation}
    \mathcal{L}_{mass} = -\frac{5}{2}g A_1^{ab}\Omega_{bc}\overline{\psi}_{\mu a}\Gamma^{\mu \nu}\psi_{\nu}^c + \frac{1}{4}gA_2^{d,abc}\Omega_{de}\overline{\chi}_{abc}\Gamma^{\mu}\psi_{\mu}^e +\frac{g}{4}\mathcal{M}^{abc}\,_{def}\overline{\chi}_{abc}\chi^{def}.
\end{equation}
where $\Gamma^{\mu}$ are the 7-dimensional Dirac Gamma matrices and
\begin{equation}
 \mathcal{M}^{abc}\,_{def} =  \frac{1}{\sqrt{2}}\delta^a_d\left[\frac{3}{32}\delta^b_e\delta^c_f B +\frac{1}{8}\delta^b_e\Omega_{fg}C^{gc} +B^{bc}\,_{ef}-C^{bc}\,_{ef}\right]
\end{equation}
obtained from (\ref{fermionbaremass}).
We can do a field redefinition for the gravitinos 
\begin{equation}
    \psi_{\mu}^a \rightarrow \psi_{\mu}^a + \frac{1}{120}\Omega^{ab}A_{1bc}^{-1}A_2^{c,efg}\Gamma_{\mu}\chi_{efg}
\end{equation}
in order to gauge away the interaction term between fermions and gravitinos, this will add a new term to the mass matrix of the fermions, (\ref{fermionbaremass}), rendering the Goldstinos null eigenvectors.
The supersymmetric AdS vacuum, (A1), is perturbatively stable, respecting the Breitenlohner--Freedman bound \cite{Breitenlohner:1982bm, Breitenlohner:1982jf, Mezincescu:1984ev}
\begin{equation}
    m^2 \ge -\frac{(d-1)^2}{4},
\end{equation}
for scalar degrees of freedom in $AdS_d$. On the other hand, (A2), is pertubatively unstable, as already found in \cite{Pernici:1984zw}, thus corroborating the hypothesis about the instability of non-supersymmetric AdS spacetimes formulated in the contest of the Swampland programme \cite{Ooguri:2016pdq}. 
For the Minkowski vacua, we obtain
\hspace{-1.2cm}
\begin{table}[H]
\rowcolors{1}{white}{gray!15}
\begin{tabular}{|c|c|}
	\hline
	$m^2_{3/2}$ & $[m^2]_{\times 2}$, $[0]_{\times 2}$  \\[2mm]
	$ m^2_{vec}$ & $[ (2 m)^2]_{\times 2}$, $[(m)^2]_{\times 2}$, $[0]_{\times 6}$ \\[2mm]
	$ m^2_{2-forms}$ &   $[0]_{\times 3}$, $[m^2]_{\times 2}$  \\[2mm]
	$m^2_{3-forms}$ &   $[m^2]_{\times 2}$, $[0]_{\times 3}$  \\[2mm]
	$m^2_{scal}$ &  $[-(4m)^2]_{\times 4}$, $[-(2m)^2]_{\times 2}$, $[0]_{\times 8}$\\[2mm]
	\hline
\end{tabular}
\centering
\caption{Masses for the Minkowski vacuum M1}
\end{table}

\hspace{-1.2cm}
\begin{table}[H]
\rowcolors{1}{white}{gray!15}
\begin{tabular}{|c|c|}
	\hline
	$m^2_{3/2}$ & $\left[\frac{1}{4}(m_1^2 -m_2^2)\right]_{\times 4}$ \\[2mm]
	$ m^2_{vec}$ & $[ (m_1)^2]_{\times 2}$, $[(m_1\pm m_2)^2]_{\times 2}$, $[0]_{\times 4}$ \\[2mm]
	$ m^2_{2-forms}$ &   $[0]_{\times 3}$, $[m_2^2]_{\times 2}$  \\[2mm]
	$m^2_{3-forms}$ &   $[m_1^2]_{\times 2}$, $[0]_{\times 3}$  \\[2mm]
	$m^2_{scal}$ &  $[0]_{\times 8}$, $[-(4 m_1)^2]_{\times 2}$, $[-(4 m_2)^2]_{\times 2}$, $[-(2m_2)^2]_{\times 2}$\\[2mm]
	\hline
\end{tabular}
\centering
\caption{Masses for the Minkowski vacuum M2}
\end{table}
In Appendix \ref{Appendix2} we present the form of the representations that make up the T-tensor, for each vacua we found. It is clear that this is not an exaustive analysis, indeed some vacua are missing from this scan, for example the Scherk--Schwarz Minkowski vacuum, with gauge group U$(1)\times\mathbb{R}^6$, first found in \cite{Samtleben:2005bp}, did not appear in our scan. In any case, it is possible to find it, once one limits the analysis to the vacua preserving a residual symmetry, in this case U$(1)$. 

\section{Outlook and Future Directions}
In section 3 we presented a new method, based on Evolutionary Strategies optimization techniques. The power of these tools combined with the thorough analysis of the numerical data allowed the reconstruction of analytical solutions and their mass spectra. The analysis has been carried out for $D=7$ space-time dimensions, for reasons already explained in the introduction, but it would be interesting to see whether our procedure can be used also for different number of dimensions and for fewer amount of supersymmetry. The number of parameters in the embedding tensor, and therefore of variables, and the number of equations grow as the number of dimensions get lower. This is illustrated in Table \ref{thetarep}, where the representations of the embedding tensor for maximal theories in each dimension are presented, giving the number of variables in the system of equations (there is the possibility to remove some of them using the H-redundancy of the scalar coset manifold). This implies that numerical methods are necessary when dealing with lower dimensional theories, because analytical results can be achieved in very special cases, either by fixing the gauging or restricting the analysis to vacua preserving some residual symmetry. Numerical methods, instead, would allow for a more general scan. On the other hand, with a large number of non-vanishing parameters in the vacua it is no longer possible to use the reconstruction techniques presented in section 3 and more subtle ways are needed if we want to obtain analytical results.
\begin{table}[H]
\rowcolors{1}{white}{gray!15}
\centering
\begin{tabular}{|c | c | c|}
\hline
D & Duality Group & $\Theta_M{}^{\alpha}$ \\ 

9 & GL(2) & $\mathbf{2}^{-3}$ + $\mathbf{3}^{+4}$  \\

8 & SL(2)$\times$ SL(3) &  $(\mathbf{2},\mathbf{3})$ + $(\mathbf{2},\mathbf{6}')$\\

7 & SL(5) & $\mathbf{15} +\mathbf{40'}$ \\

6 & SO(5,5) & $\mathbf{144_c}$\\

5 & $E_{6(6)}$ & $\mathbf{351'}$\\

4 & $E_{7(7)}$ &  $\mathbf{912}$ \\

3 & $E_{8(8)}$ & $\mathbf{1}+\mathbf{3875}$\\
\hline
\end{tabular}
\caption{Representations of the Embedding Tensor}
\label{thetarep}
\end{table}
We also compared our method to some optimization tools available from TensorFlow libraries (Adam optimizer) and to other algorithms as well, finding CMA-ES and our implementations more efficient, given the cospicuous extension of the literature in Optimization it has been impossible to test our problem against many of the optimization techniques. Another interesting step for the future is to compare our methodology with other numerical approaches; maximal supergravities in $D=7$ space-time dimensions are a good testbed due to  the small number of variables and the interesting physical features. 
\pagebreak
\appendix
\section{Covariance Matrix Adaptation - Evolutionary \\ Strategy (CMA-ES)}
\label{sec:CMA}
In this Appendix, we present some details about the CMA-ES algorithm and its operational mode.
CMA-ES, in short, CMA, is an evolution strategy (ES) algorithm that, as in the case of GA, only needs the fitness function as accessible information \cite{6790628, 6790790, 542381, appppppp, igel2007covariance, hansen2001completely, hansen2004evaluating, ros2008simple, Hansen06thecma}. Therefore, differently from GD or stocastic GD, we do not require the function to be differentiable, it can also be not continuous. CMA, as the name suggests, is based on the ``adaption'' of a normal distribution to the fitness function under consideration (with its level curves). Let us analyse the algorithm in depth and consider a multivariate normal distribution, $\mathcal{N}(\mathbf{m}, \mathbf{C})$, which is determined by the mean $\mathbf{m} \in \mathbb{R}^n$ (in our case, n is the number of free parameters in the embedding tensor) and by the symmetric, positive definite covariance matrix $\mathbf{C}\in \mathbb{R}^{n\times n}$. Covariance matrices are associated with the ellipsoid $\{\mathbf{x}\in \mathbb{R}^n | \mathbf{x}^T \mathbf{C}^{-1} \mathbf{x}=1\}$, where the principal axis of the latter are the eigenvectors of $\mathbf{C}$, and the lengths of the squared axis are the eigenvalues of the covariance matrix.
We can always diagonalize the covariance matrix by means of an orthogonal matrix $\mathbf{B}$ whose columns are the eigenvectors of $\mathbf{C}$ with unit length, $\mathbf{C}= \mathbf{B}(\mathbf{D})^2\mathbf{B}^T$. Then, it is also possible to write the normal distribution as 
\begin{equation}
    \mathcal{N}(\mathbf{m}, \mathbf{C}) \sim \mathbf{m} +\mathcal{N}(\mathbf{0}, \mathbf{C})\sim \mathbf{m} + \mathbf{C}^{-1/2}\mathcal{N}(\mathbf{0}, \mathbf{I}) \sim \mathbf{m} + \mathbf{BDB}^T\mathcal{N}(\mathbf{0},\mathbf{I}),
\end{equation}
with $\mathbf{I}$ the $n\times n$ identity matrix. 
At each step of the process, we generate a new population of points (which in GA are called offsprings) by drawing them from a multivariate normal distribution:
\begin{equation}
    \mathbf{x}_j^{g+1} \sim \mathcal{N}\big(\mathbf{m}^g, (\sigma^g)^2 \mathbf{C}^g\big) \hspace{1cm} \text{with} \hspace{0.3cm} j=1,...,\lambda
\end{equation}
Superscripts $g$, $g+1$, etc. label the generations, $\lambda$ is the population size, and $\sigma^g \in \mathbb{R}^+$ is the ``overall'' standard deviation (step size) at generation $g$. Now we need to explain how the mean, the covariance matrix, and the standard deviation are computed for the next generation $g+1$.

\subsubsection*{The mean}
The new mean $\mathbf{m}^{g+1}$ is simply selected with a weighted average of the $\mu$ best points of the population:
\begin{equation}
    \mathbf{m}^{g+1} = \sum_{j=1}^{\mu} w_j \mathbf{x}_{j:\lambda}^{g+1}, \hspace{1cm} \text{with} \hspace{0.3cm} \sum_{j=1}^{\mu}w_j = 1 \hspace{0.3cm} \text{and} \hspace{0.3cm}w_j >0.
    \label{meanCMA}
\end{equation}
$w_j \in \mathbb{R}^+$ with $j=1,..., \mu$ are positive ordered weights, that is, $w_1 \ge w_2 \ge ... \ge w_{\mu} >0$. If $w_j = 1/\mu$ for each j, we obtain the mean value for the best $\mu$ points. $\mathbf{x}_{j:\lambda}^{g+1}$ represents the j-th best individual of the population, meaning, $f(\mathbf{x}_{1:\lambda}^{g+1})\le f(\mathbf{x}_{2:\lambda}^{g+1})\le ... \le f(\mathbf{x}_{\lambda:\lambda}^{g+1})$. An essential quantity is the \textbf{variance effective selection mass} 
\begin{equation}
    \mu_{eff} = \big(\sum_{j=1}^{\mu}w_j^2\big)^{-1}.
\end{equation}
It is possible to show, from the definition of $w_j$ that $1\le \mu_{eff} \le \mu$ and that $\mu_{eff}=\mu$ only in the case where all the weights are the same and equal to $1/\mu$. Usually, $\mu \thickapprox \lambda/2$ and $w_i \propto \mu -i +1$.
\subsubsection*{The covariance matrix}
First, let us define the empirical covariance matrix $\mathbf{C}_{emp}^{g+1}$, which is nothing more than an estimate of the covariance matrix $\mathbf{C}^g$:
\begin{equation}
    \mathbf{C}_{emp}^{g+1} = \frac{1}{\lambda -1} \sum_{i=1}^{\lambda}\bigg(\mathbf{x}_i^{g+1} -\frac{1}{\lambda}\sum_{j=1}^{\lambda}\mathbf{x}_j^{g+1}\bigg)\bigg(\mathbf{x}_i^{g+1} -\frac{1}{\lambda}\sum_{j=1}^{\lambda}\mathbf{x}_j^{g+1}\bigg)^T.
\end{equation}
This estimator has to be modified to obtain a maximum likelihood estimator of $\mathbf{C}^g$, by defining
\begin{equation}
    \mathbf{C}_{\lambda}^{g+1} = \frac{1}{\lambda}\sum_{j=1}^{\lambda}\big(\mathbf{x}_j^{g+1}-\mathbf{m}^g\big)\big(\mathbf{x}_j^{g+1}-\mathbf{m}^g\big)^T.
\end{equation}
The difference between $\mathbf{C}_{emp}^{g+1}$ and $\mathbf{C}_{\lambda}^{g+1}$ is what is used as the mean value. The former uses the mean obtained from the entire population, thus estimating the variance of the sampled points, and the latter instead uses the mean obtained by (\ref{meanCMA}), therefore estimating the sampled steps, $\mathbf{x}_j^{g+1} - \mathbf{m}^g$. We are going to modify this estimator again and define
\begin{equation}
    C_{\mu}^{g+1} = \sum_{j=1}^{\mu} w_j \left(\mathbf{x}_{j:\lambda}^{g+1}-\mathbf{m}^g\right)\left(\mathbf{x}_{j:\lambda}^{g+1}-\mathbf{m}^g\right)^T.
\end{equation}
$\mathbf{C}_{\mu}^{g+1}$ is an estimator of the variance of selected steps (the best / successful steps $\mu$). We have some conditions on $\mu_{eff}$ in order for $\mathbf{C}_{\mu}^{g+1}$ to be a \textbf{reliable} estimator. Indeed, $\mu_{eff}$ has to be large enough to prevent the condition numbers (given a matrix A and a linear system Ax=b, with x unknown, they measure how sensitive the solution of the system to a change in b is, high condition numbers imply that small changes in b generate huge modifications in the solution) of $\mathbf{C}_{\mu}^{g+1}$ to be smaller than 10 for the fitness function of the sphere: $f_{sphere}(\mathbf{x})= \sum_{i=1}^n x_i^2$; empirically, it is seen that $\mu_{eff}\thickapprox 10n$ is a good choice. To avoid this problem for a small population, we will modify the update of the covariance matrix again. In order to obtain an algorithm that converges faster, we need a small population, on the other hand, to obtain a more global search the population has to increase. For a small population, also $\mu_{eff}\thickapprox \lambda/4$ (which is the choice to take to have reasonable $w_j$) has to be small, then $\mathbf{C}_{\mu}^{g+1}$ is not a reliable estimator, in order to circumvent this, we define a new covariance matrix that takes into consideration the information we have from previous generations. Defining $\mathbf{C}^0 = \mathbf{I}$ and the learning rate $0 < c_{cov} \le 1$, then 
\begin{align}
\begin{split}
    \mathbf{C}^{g+1} =& (1-c_{cov})\mathbf{C}^g +c_{cov}\bigg(\frac{1}{\sigma^{g}}\bigg)^2 \mathbf{C}_{\mu}^{g+1}\\ =& (1-c_{cov})\mathbf{C}^g + c_{cov}\sum_{j=1}^{\mu} w_j\left(\frac{\mathbf{x}_{j:\lambda}^{g+1}-\mathbf{m}^g}{\sigma^g}\right)\left(\frac{\mathbf{x}_{j:\lambda}^{g+1}-\mathbf{m}^g}{\sigma^g}\right)^T.
\end{split}
\label{rankmu}
\end{align}
Step sizes $\sigma^g$ have been integrated to ensure that $\mathbf{C}_{\mu}^g$ from different generations are comparable. If $c_{cov}=1$ the covariance matrix collapses to $\mathbf{C}_{\mu}^{g+1}$ and no information from previous generations is retained, on the other hand, if $c_{cov}=0$, $\mathbf{C}^{g+1} = \mathbf{C}^0$ and there is no learning. This kind of update, represented in (\ref{rankmu}) update, is called rank-$\mu$ update, because the sum goes from 1 to $\mu$. Eq. (\ref{rankmu}) is iterative and can be expanded as
\begin{equation}
    C^{g+1} = (1-c_{cov})^{g+1}\mathbf{C}^0 +c_{cov}\sum_{j=0}^g (1-c_{cov})^{g-j}\bigg(\frac{1}{\sigma^j}\bigg)^2 \mathbf{C}_{\mu}^{j+1}.
\end{equation}
Picking high values for $c_{cov}$ leads to degenerate covariance matrices, while small values imply slow learning, a good choice is $c_{cov}\thickapprox \mu_{eff}/n^2$. Small population sizes $\lambda$ lead to a large number of generations and therefore to a faster adaptation for the covariance matrix. 
A final step is necessary to update the covariance matrix: \textbf{cumulation}. In fact, information about ``signs'' of the steps the strategy took generation after generation has not been used so far. To do so, we introduce the \textbf{evolution path}. An evolution path is any sequence of succesive steps taken by the strategy, taking the sum of these steps is referred as cumulation; for instance, for three steps we have
\begin{equation}
    \frac{\mathbf{m}^{g+1} - \mathbf{m}^g}{\sigma^g}+\frac{\mathbf{m}^{g} - \mathbf{m}^{g-1}}{\sigma^{g-1}}+\frac{\mathbf{m}^{g-1} - \mathbf{m}^{g-2}}{\sigma^{g-2}}.
\end{equation} 
Defining the 0-th order evolution path $\mathbf{p}_c^0 = \mathbf{0}$, we use the exponential smoothing and define iteratively
\begin{equation}
    \mathbf{p}_c^{g+1}=(1-c_c)\mathbf{p}_c^g +\sqrt{c_c(2-c_c)\mu_{eff}}\frac{\mathbf{m}^{g+1}-\mathbf{m}^g}{\sigma^g},
\end{equation}
with $0 \ge c_c \le 1$ a new learning rate for the evolution path, the normalisation factor \\ $\sqrt{c_c (2-c_c)\mu_{eff}}$ is dictated by the demand that $\mathbf{p}_c^{g+1}$ be extracted from a normal distribution $\mathcal{N}(\mathbf{0}, \mathbf{C})$. When $c_c=0$ there is no learning and $\mathbf{p}_c^{g}=\mathbf{0}$. Putting everything together, we obtain the update of the covariance matrix:
\begin{align}
    \begin{split}
    \mathbf{C}^{g+1} =& (1 - c_{cov})\mathbf{C}^g +\frac{c_{cov}}{\mu_{cov}}\mathbf{p}_c^{g+1}\mathbf{p}_c^{g+1\,T}\\ &+ c_{cov}\left(1-\frac{1}{\mu_{cov}}\right)
    \sum_{j=1}^{\mu}w_j \left(\frac{\mathbf{x}_{j:\lambda}^{g+1}-\mathbf{m}^g}{\sigma^g}\right)\left(\frac{\mathbf{x}_{j:\lambda}^{g+1}-\mathbf{m}^g}{\sigma^g}\right)^T,
    \end{split}
    \label{covmatCMA}
\end{align}
with $\mu_{cov}\ge 1$ and should be $\mu_{cov}=\mu_{eff}$. Eq. (\ref{covmatCMA}) reduces to eq. (\ref{rankmu}) in the case $\mu_{cov}\rightarrow \infty$, so information from the last generation is taken into consideration by the rank-$\mu$ update and information from previous generations, instead, is exploited by the evolution path update, which is relevant above all for small population sizes.
\subsubsection*{The step size}
An evolution path is also used to update the step size $\sigma$ with a method called \textbf{cumulative step size adaptation}:
\begin{itemize}
    \item Whenever the evolution path is long, the steps are going in the same direction (approximatively), so they are correlated. Consequently, we can cover the same distance with longer but fewer steps, and the step size must be increased.
    \item When the evolution path is short, the steps cancel among each other, and the step size should be decreased.
    \item The optimal situation is that the steps are totally uncorrelated and orthogonal with respect to the previous and following ones.
\end{itemize}
We need to define what long- and short-evolution paths mean. In this respect, we compare the latter with the expected length under random selection, which means that the steps are not correlated with each other. If our strategy finds that the evolution paths are longer than the uncorrelated ones, $\sigma$ has to be increased, and vice versa. \\ The evolution path $\mathbf{p}_c^{g+1}$ depends on its direction, therefore we define the conjugate path
\begin{equation}
    \mathbf{p}_{\sigma}^{g+1} = (1- c_{\sigma})\mathbf{p}_{\sigma}^g + \sqrt{c_{\sigma}(2-c_{\sigma})\mu_{eff}}(\mathbf{C}^{g})^{-\frac{1}{2}}\frac{\mathbf{m}^{g+1}-\mathbf{m}^g}{\sigma^g},
    \label{stepsizeevolution}
\end{equation}
with $0<c_{\sigma}<1$ a learning rate and $(\mathbf{C}^{g})^{-\frac{1}{2}} \equiv \mathbf{B}^g (\mathbf{D}^{g})^{-1}\mathbf{B}^{g\, T}$. Whenever $(\mathbf{C}^{g})^{-\frac{1}{2}}\neq \mathbf{I}$ it aligns step $\mathbf{m}^{g+1} - \mathbf{m}^g$ with the coordinate system produced by $\mathbf{B}^g$. In particular, $\mathbf{B}^{g \, T}$ rotates the system so that the columns of $\mathbf{B}^g$ become the axis. $(\mathbf{D}^{g})^{-1}$ rescales the length of the axis so that they measure the distances in the same way. $\mathbf{B}^g$ rotates everything back, allowing us to compare the directions of the various steps. By adding the matrix $(\mathbf{C}^{g})^{-\frac{1}{2}}$ in eq.(\ref{stepsizeevolution}) we ensure the independence of $\mathbf{p}_{\sigma}^{g+1}$ from the direction of the steps. Then we compare the length of $\mathbf{p}_{\sigma}^{g+1}$ with the expected length of the evolution path obtained from random selection $E\left[||\mathcal{N}(\mathbf{0},\mathbf{I})||\right]$ and define the step size 
\begin{equation}
\label{sigmaCMA}
    \sigma^{g+1} = \sigma^g \exp \left(\frac{c_{\sigma}}{d_{\sigma}}\left(\frac{||\mathbf{p}_{\sigma}^{g+1}||}{E\left[||\mathcal{N}(\mathbf{0},\mathbf{I})||\right]}-1\right)\right),
\end{equation}

where $d_{\sigma}\thickapprox 1$ is a damping parameter and $E\left[||\mathcal{N}(\mathbf{0}, \mathbf{I})||\right] = \sqrt{2}\Gamma\left(\frac{n+1}{2}\right)/\Gamma\left(\frac{n}{2}\right)\thickapprox \sqrt{n}+\mathcal{O}(1/n)$ is the expectation value of the Euclidean norm for a multivariate normal distribution with the identity matrix as the covariance matrix. From eq. (\ref{sigmaCMA}) we can see that, whenever $||\mathbf{p}_{\sigma}^{g+1}||>E\left[||\mathcal{N}(\mathbf{0}, \mathbf{I})||\right]$, $\sigma^{g}$ increases and viceversa when $||\mathbf{p}_{\sigma}^{g+1}||<E\left[||\mathcal{N}(\mathbf{0}, \mathbf{I})||\right]$.\\  It has been proved, in a survey about Black-Box optimizations \cite{hansen:hal-00545727}, that CMA-ES outranks other 31 optimisation algorithms and that its performance is outstanding for rugged and ill-conditioned functions with large search spaces.

\section{T-tensor at the Vacuum in the 7 dimensional \\ Theory}
\label{Appendix2}
In this Appendix we provide an instance of the value of the irreducible USp$(4)$ representations composing the T-tensor generating the vacua of Table 4.22. For all the examples, we have chosen a basis where
\begin{equation}
    \Omega = \mathbb{1} \otimes i\sigma_2 .
\end{equation}
\\
\textbf{A1}. The maximal supersymmetric Anti de Sitter vacuum has 
\begin{equation}
    Bs= \kappa, \hspace{1cm} B^{ab}{}_{cd} = C^{ab}{}_{cd} = C^{ab} = \mathbf{0}.
\end{equation}
\textbf{A2}. The SO(4) non-supersymmetric AdS vacuum is given by the following representations:
\begin{equation}
    Bs=\kappa, \hspace{1cm}C^{ab}{}_{cd} = C^{ab} = \mathbf{0},
\end{equation}
\begin{align}
    \begin{split}
    B^{12}{}_{12}&=B^{12}{}_{43}=B^{34}{}_{21}=B^{34}{}_{34}=\frac{\kappa}{6},\\
    B^{13}{}_{31}&=B^{14}{}_{41}=B^{23}{}_{32}=B^{24}{}_{42}= \frac{\kappa}{12}.
    \end{split}
\end{align}
There are other non-vanishing entries in the $\mathbf{14}$ representation, which are related by symmetries in the indices and therefore have not been reported here.
\pagebreak

\textbf{M1}.
\begin{equation}
    Bs= \kappa, \hspace{1cm} C^{12}=C^{43}=\frac{5}{4}\kappa,
\end{equation}
\begin{align}
    B^{12}{}_{21}&= B^{12}{}_{34}=B^{34}{}_{12}=B^{34}{}_{43}=\frac{1}{4}\kappa,\\
    B^{13}{}_{13}&=B^{14}{}_{14}=B^{23}{}_{23}=B^{24}{}_{24}=\frac{1}{8}\kappa,\\
    B^{13}{}_{41}&=B^{14}{}_{13}=B^{23}{}_{24}=B^{24}{}_{32}=\frac{5i}{8}\kappa,
\end{align}
\begin{align}
    C^{14}{}_{13}=C^{23}{}_{24}=C^{31}{}_{14}=C^{42}{}_{23}=\frac{5i}{8}\kappa.
\end{align}
Here, one needs always to keep in mind that there are other, non-reported, entries of these tensors which are related to the ones above by symmetries of representations.

\textbf{M2}.
\begin{equation}
    Bs=\kappa_1, \hspace{1cm} C^{14}=C^{23}=4i\kappa_2,
\end{equation}
\begin{align}
    B^{12}{}_{21}&= B^{12}{}_{34}=B^{34}{}_{12}=B^{34}{}_{43}=\frac{1}{4}\kappa_1,\\
    B^{13}{}_{13}&=B^{24}{}_{24}=\frac{3}{4}\kappa_1,\\
    B^{14}{}_{41}&=B^{23}{}_{32}=\frac{\kappa_1}{2},
\end{align}
\begin{align}
    C^{14}{}_{12}&=C^{21}{}_{14}=C^{21}{}_{23}=C^{23}{}_{43}=C^{32}{}_{12}=C^{34}{}_{14}=C^{34}{}_{23}=C^{41}{}_{34}=i\kappa_2,\\
    C^{13}{}_{11}&=C^{13}{}_{33}=C^{42}{}_{22}=C^{42}{}_{44}=2i\kappa_2.
\end{align}
Again, symmetries must be imposed on these tensors.
%
\bigskip
\section*{Acknowledgments}

\noindent We would like to thank H.~Samtleben for useful correspondence and S.~Giardino and G.~Inverso for helpful discussions. Furthermore, we are really grateful to G.~Dall'Agata for the precious suggestions and comments.

%

\bibliography{Biblio}
\bibliographystyle{unsrt}

\end{document}